\documentclass[9pt,twocolumn,twoside]{pnas-new}
  
  \usepackage{amsmath}
  \usepackage{amsfonts}
  \usepackage{gensymb}
  \usepackage{mathrsfs,amsmath} 
  \usepackage{amssymb}
  \usepackage{graphicx}
  \usepackage{siunitx}
  \usepackage{array,booktabs}
  \newcolumntype{M}{>{$} c <{$}}
  
  \usepackage[]{units}
  \usepackage{IEEEtrantools}
  \usepackage{mathtools}
  \usepackage{amsfonts}
  \usepackage[]{xcolor}
  \usepackage{array}
  \usepackage{siunitx,booktabs}
  \usepackage{multirow}

  \usepackage{psfrag}
  
  \usepackage{etoolbox}
  \appto\normalsize{\belowdisplayshortskip=\belowdisplayskip}
  \appto\small{\belowdisplayshortskip=\belowdisplayskip}
  \appto\footnotesize{\belowdisplayshortskip=\belowdisplayskip}
  \setcitestyle{round}
  
  \templatetype{pnasresearcharticle} 

\title{Lipid bilayer mediates ion-channel cooperativity in a model of hair-cell mechanotransduction}

  \author[a]{Francesco Gianoli} 
  \author[b,c,1,2]{Thomas Risler} 
  \author[a,1,2]{Andrei S.~Kozlov}
  
  \affil[a]{Department of Bioengineering, Imperial College London, London SW7 2AZ, UK}
  \affil[b]{Laboratoire Physico Chimie Curie, Institut Curie, PSL Research University, CNRS, 26 rue d'Ulm, 75005 Paris, France}
  \affil[c]{Sorbonne Universit\'es, UPMC Univ Paris 06, CNRS, Laboratoire Physico Chimie Curie, 75005 Paris, France}
  
  \leadauthor{Gianoli} 
  
\significancestatement{
Hearing relies on molecular machinery that consists of springs stretched by mechanical stimuli and mechanosensitive ion channels responding to the generated tension. Reproducing the experimental data theoretically without requiring unrealistically large conformational changes of the channels has been a longstanding hurdle. Here, we propose and develop a model with two mobile channels per spring, coupled by elastic forces within the membrane. The relative motion of the channels following their cooperative opening and closing produces the required change in spring extension. This study lies at the interface between the fields of membrane mechanics and mechanotransduction in the inner ear. It describes a physiological function for the bilayer-mediated cooperativity between mechanosensitive ion channels in a vertebrate sensory system.
}

\authorcontributions{A.S.K. conceived the proposed mechanism; A.S.K. initiated and supervised the project; F.G., T.R., and A.S.K. developed the model; T.R. supervised the work on the theoretical formalism; F.G. and T.R. implemented the model in a computer program; F.G., T.R., and A.S.K. generated and analyzed the results; F.G. produced the figures; F.G., T.R., and A.S.K. wrote the paper.} 
  
\authordeclaration{The authors declare no conflict of interest.}
\equalauthors{\textsuperscript{1}T.R. and A.S.K. contributed equally to this work.}
\correspondingauthor{\textsuperscript{2}To whom correspondence should be addressed. E-mails: a.kozlov@imperial.ac.uk or thomas.risler@curie.fr.}


\begin{abstract}
Mechanoelectrical transduction in the inner ear is a biophysical process underlying the senses of hearing and balance. The key players involved in this process are mechanosensitive ion channels. They are located in the stereocilia of hair cells and opened by the tension in specialized molecular springs, the tip links, connecting adjacent stereocilia. When channels open, the tip links relax, reducing the hair-bundle stiffness. This gating compliance makes hair cells especially sensitive to small stimuli. The classical explanation for the gating compliance is that the conformational rearrangement of a single channel directly shortens the tip link. However, to reconcile theoretical models based on this mechanism with experimental data, an unrealistically large structural change of the channel is required. Experimental evidence indicates that each tip link is a dimeric molecule, associated on average with two channels at its lower end. It also indicates that the lipid bilayer modulates channel gating, although it is not clear how. Here, we design and analyze a model of mechanotransduction where each tip link attaches to two channels, mobile within the membrane. Their states and positions are coupled by membrane-mediated elastic forces arising from the interaction between the channels’ hydrophobic cores and that of the lipid bilayer. This coupling induces cooperative opening and closing of the channels. The model reproduces the main properties of hair-cell mechanotransduction using only realistic parameters constrained by experimental evidence. This work provides an insight into the fundamental role that membrane-mediated ion-channel cooperativity can play in sensory physiology.
\end{abstract}

\dates{This manuscript was compiled on \today}
\doi{\url{www.pnas.org/cgi/doi/10.1073/pnas.XXXXXXXXXX}}
  
\begin{document}
  
  
  \maketitle
  
  \thispagestyle{firststyle}
  \ifthenelse{\boolean{shortarticle}}{\ifthenelse{\boolean{singlecolumn}}{\abscontentformatted}{\abscontent}}{}

\dropcap{M}echanoelectrical transduction (MET) in the inner ear occurs when mechanical forces deflect the stereocilia of hair cells, changing the open probability of mechanosensitive ion channels located in the stereociliary membrane~\cite{hudspeth_how_1989,peng_integrating_2011,fettiplace_physiology_2014}. Channel gating (opening and closing) and stereocilia motion are directly coupled by tip links, extracellular filaments that connect the tip of each stereocilium to the side of its taller neighbor~\cite{pickles_cross-links_1984}. Tip links act as molecular springs, whose tension determines the channels’ open probability. Reciprocally, channel gating affects tension in the tip links and consequently the stiffness of the whole hair bundle. This phenomenon, known as gating compliance~\cite{howard_compliance_1988}, is a key feature of hair-cell mechanics and contributes to the auditory system's high sensitivity and sharp frequency tuning~\cite{fettiplace_physiology_2014, hudspeth_putting_2000}. Its mechanism, however, remains unclear. The classical model of mechanotransduction ascribes gating compliance to the gating swing, a change in the extension of the tip link due to the conformational rearrangement of a single MET channel upon gating~\cite{howard_compliance_1988, markin_gating-spring_1995}. To reproduce experimental data theoretically, however, the amplitude of the gating swing must be comparable to, or even greater than, the size of a typical ion channel~\cite{martin_negative_2000,van_netten_channel_2003,tinevez_unifying_2007,sul_gating_2010}. This requirement constitutes an issue that is often acknowledged~\cite{peng_integrating_2011, fettiplace_physiology_2014, albert_comparative_2016} but still unresolved.
\begin{figure*}[h]
\centering
\includegraphics[width=\textwidth]{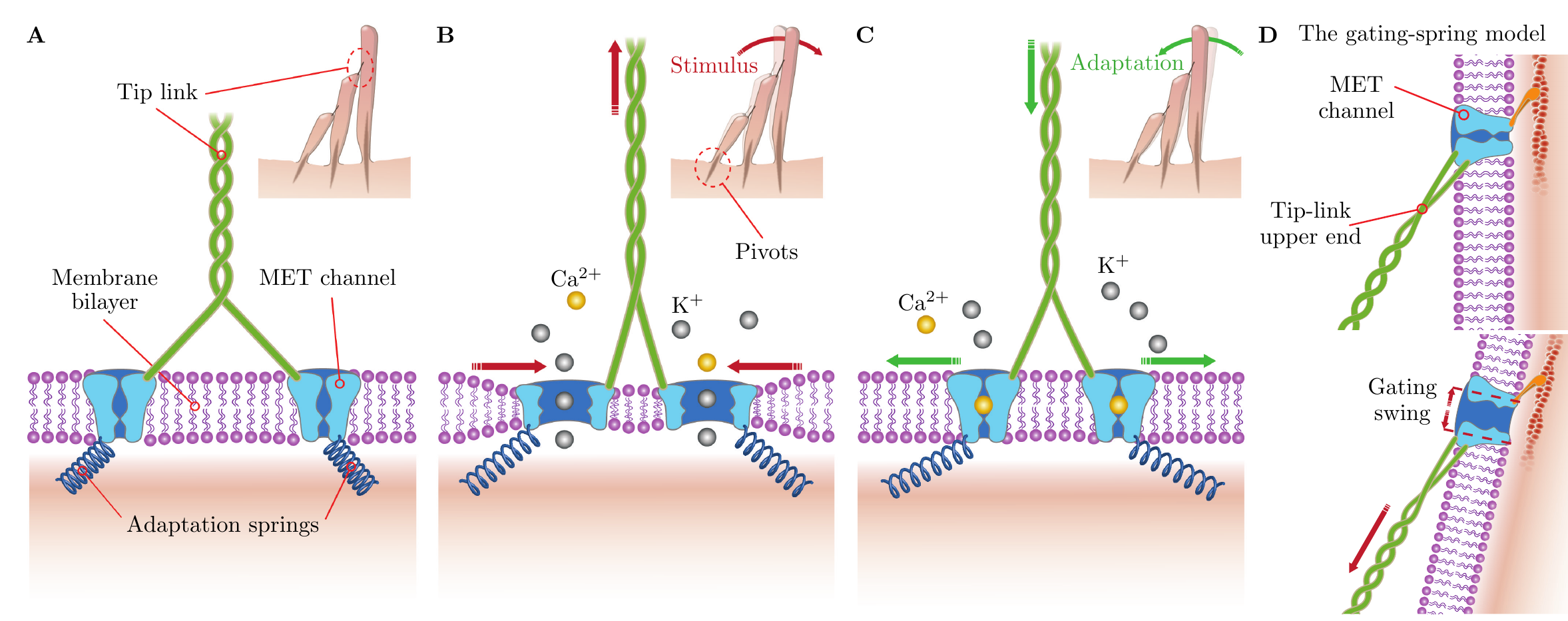}
\caption{Illustration of the model and its main features. (\textit{A--C}) \textit{Insets} show a side view of a typical mammalian hair bundle with three rows of stereocilia, which taper at their basal insertion points where a pivoting stiffness maintains them upright (``pivots''). The direction of mechanosensitivity is from left to right, with positive displacements to the right. It corresponds to the `$X$' axis as defined later in the text. The main images show an enlarged view of the lower end of a single tip link, connected to two MET channels within the lipid bilayer. The channels are linked to the cell cytoskeleton via two adaptation springs. Three configurations are shown: in the absence of a stimulus (\textit{A}), when a positive stimulus is applied (\textit{B}), and when fast adaptation takes place (\textit{C}). (\textit{D}) Closed (\textit{Upper}) and open (\textit{Lower}) configurations of the classical gating-spring model for comparison. A single mechanotransduction channel is located at the tip link's upper end. It is firmly anchored to the cytoskeleton and unable to change its position at the short timescale of channel gating. The gating swing is the amplitude of the channel's conformational change along the tip link's axis that relaxes the tip link when the channel opens.
}\label{FigTwoChannelModel}
\end{figure*}

The classical model posits a single MET channel connected to the tip link’s upper end, near myosin motors that regulate tip-link tension~\cite{hudspeth_making_2008}. Electrophysiological recordings, however, point to two channels per tip link~\cite{denk_calcium_1995,ricci_tonotopic_2003,beurg_large-conductance_2006, beurg_localization_2009}, which is in accord with its dimeric structure \cite{kazmierczak_cadherin_2007, kachar_high-resolution_2000}. Furthermore, high-speed Ca\textsuperscript{2+} imaging shows that the channels are located at the tip link’s lower end~\cite{beurg_localization_2009}. This result has been corroborated by the expression patterns of key mechanotransduction proteins within the hair bundle, which interact with protocadherin-15, the protein constituting the lower end of the tip link (reviewed in ref.~\cite{zhao_elusive_2015}). Together, these findings turned textbook views of molecular mechanotransduction in the inner ear literally upside-down~\cite{kandel_principles_2013, spinelli_bottoms_2009}. Moreover, experimental data suggest that the lipid bilayer surrounding the channels modulates their open probability as well as the rates of slow and fast adaptation~\cite{hirono_hair_2004, peng_adaptation_2016}, although it is not clear how.

In this work, we propose and explore a quantitative model of hair-cell mechanotransduction that incorporates the main pieces of evidence accumulated since the publication of the classical gating-spring model some 30 y ago \cite{howard_compliance_1988}. Our proposal relies on the cooperative gating of two MET channels per tip link, which are mobile within the membrane and coupled by elastic forces mediated by the lipid bilayer. The model accounts for the number and location of MET channels and reproduces the observed hair-cell mechanics quantitatively, using only realistic parameters. Furthermore, it provides a framework that can help understand some as-yet-unexplained features of hair-cell mechanotransduction.

\section*{Results}

\subsection*{Model Description}

\begin{figure*}[h]
\centering
\includegraphics[width=0.9\textwidth]{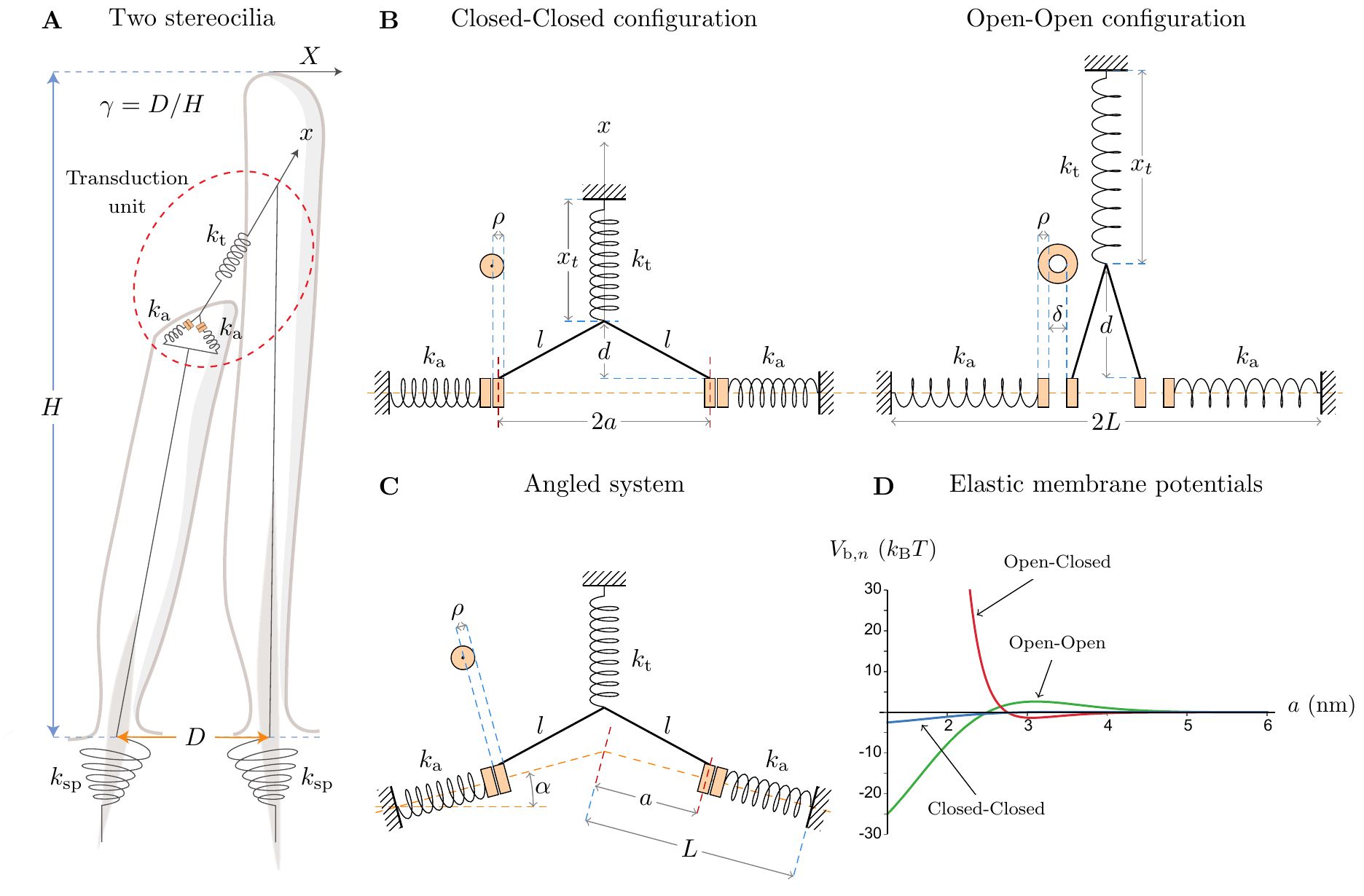}
\caption{Schematic representation of the model with its main geometrical parameters. (\textit{A}) Two adjacent stereocilia are represented with their basal elastic linkages to the cuticular plate. (\textit{B} and \textit{C}) The transduction unit, encircled in dashed red, is detailed for two different geometries: when the tip link’s central axis is perpendicular to the stereociliary membrane (\textit{B}) and in the generic case where the stereociliary membrane makes an angle with the perpendicular to the tip link’s central axis (\textit{C}). In \textit{B}, two different realizations are displayed: when tension in the tip link is low, in which case the channels are most likely to be closed (\textit{B, Left}), and when tension in the tip link is high, in which case the channels are most likely to be open (\textit{B, Right}). In \textit{C}, only the case of low tip-link tension is represented. (\textit{D}) We plot here the elastic membrane potentials in units of $k_{\rm B}T$ and as functions of the distance $a$ between the channels and the tip link's central axis. Each curve corresponds to a different configuration of the channel pair: CC (blue), OC (red), and OO (green). The analytic expressions of these potentials, together with the values of the associated parameters, are given in \textit{Materials and Methods}.
}\label{FigSchematicModel}
\end{figure*}

We describe here the basic principles of our model, illustrated in Fig.~\ref{FigTwoChannelModel}. Structural data indicate that the tip link is a dimeric, string-like protein that branches at its lower end into two single strands, which anchor to the top of the shorter stereocilium~\cite{kachar_high-resolution_2000,kazmierczak_cadherin_2007}. The model relies on three main hypotheses. First, each strand of the tip link connects to one MET channel, mobile within the membrane. Second, an intracellular spring---referred to as the adaptation spring---anchors each channel to the cytoskeleton, in agreement with the published literature~\cite{fettiplace_physiology_2014, howard_hypothesis:_2004, zhang_ankyrin_2015, powers_stereocilia_2012}. Third, and most importantly, the two MET channels interact via membrane-mediated elastic forces, which are generated by the mismatch between the thickness of the hydrophobic core of the bare bilayer and that of each channel~\cite{nielsen_energetics_1998}. Such interactions have been observed in a variety of transmembrane proteins, including the bacterial mechanosensitive channels of large conductance (MscL)~\cite{wiggins_analytic_2004, ursell_cooperative_2007, phillips_emerging_2009, grage_bilayer-mediated_2011, haselwandter_connection_2013}. Since the thickness of the channel’s hydrophobic region changes during gating, this hydrophobic mismatch induces a local deformation of the membrane that depends on the channel’s state~\cite{wiggins_analytic_2004, ursell_cooperative_2007, phillips_emerging_2009, haselwandter_connection_2013}. For a closed channel, the hydrophobic mismatch is small, and the membrane is barely deformed. An open channel's hydrophobic core, however, is substantially thinner, and the bilayer deforms accordingly~\cite{ursell_cooperative_2007, phillips_emerging_2009}. When the two channels are sufficiently near each other, the respective bilayer deformations overlap, and the overall membrane shape depends both on the states of the channels as well as on the distance between them. As a result, the pair of MET channels is subjected to one of three different energy landscapes: open--open (OO), open--closed (OC), or closed--closed (CC)~\cite{ursell_cooperative_2007}. The effects of this membrane-mediated interaction are most apparent at short distances: The potentials strongly disfavor the OC state, favor the OO state, and generate an attractive force between the two channels when they are both open.

Channel motion as a function of the imposed external force can be pictured as follows (Fig.~\ref{FigTwoChannelModel} and Movie~S1). When tip-link tension is low, the two channels are most likely to be closed, and they are kept apart by the adaptation springs; at this large inter-channel distance, the membrane-mediated interaction between them is negligible (Fig.~\ref{FigTwoChannelModel}\textit{A}). When a positive deflection is applied to the hair bundle, tension in the tip link rises. Consequently, the channels move toward one another and their open probabilities increase (Fig.~\ref{FigTwoChannelModel}\textit{B}). When the inter-channel distance is sufficiently small, the membrane's elastic energy favors the OO state, and both channels open cooperatively. As a result, the attractive membrane interaction in the OO state enhances their motion toward one another (red horizontal arrows, Fig.~\ref{FigTwoChannelModel}\textit{B} and Movie~S1), which provides an effective gating swing that is larger than the conformational change of a single channel (red vertical arrow, Fig.~\ref{FigTwoChannelModel}\textit{B}). Eventually, the channels close---for example due to Ca\textsuperscript{2+} binding~\cite{choe_model_1998, cheung_ca2+_2006}---and the membrane-mediated interactions become negligible (Fig.~\ref{FigTwoChannelModel}\textit{C}). Now the adaptation springs can pull the channels apart. Their lateral movement away from each other increases tip-link tension and produces the twitch, a hair-bundle movement associated with fast adaptation~\cite{cheung_ca2+_2006, benser_rapid_1996, ricci_active_2000}.
  
\begin{table*}[h]
\centering
\caption{Parameters of the model}
\includegraphics[width=0.95\textwidth]{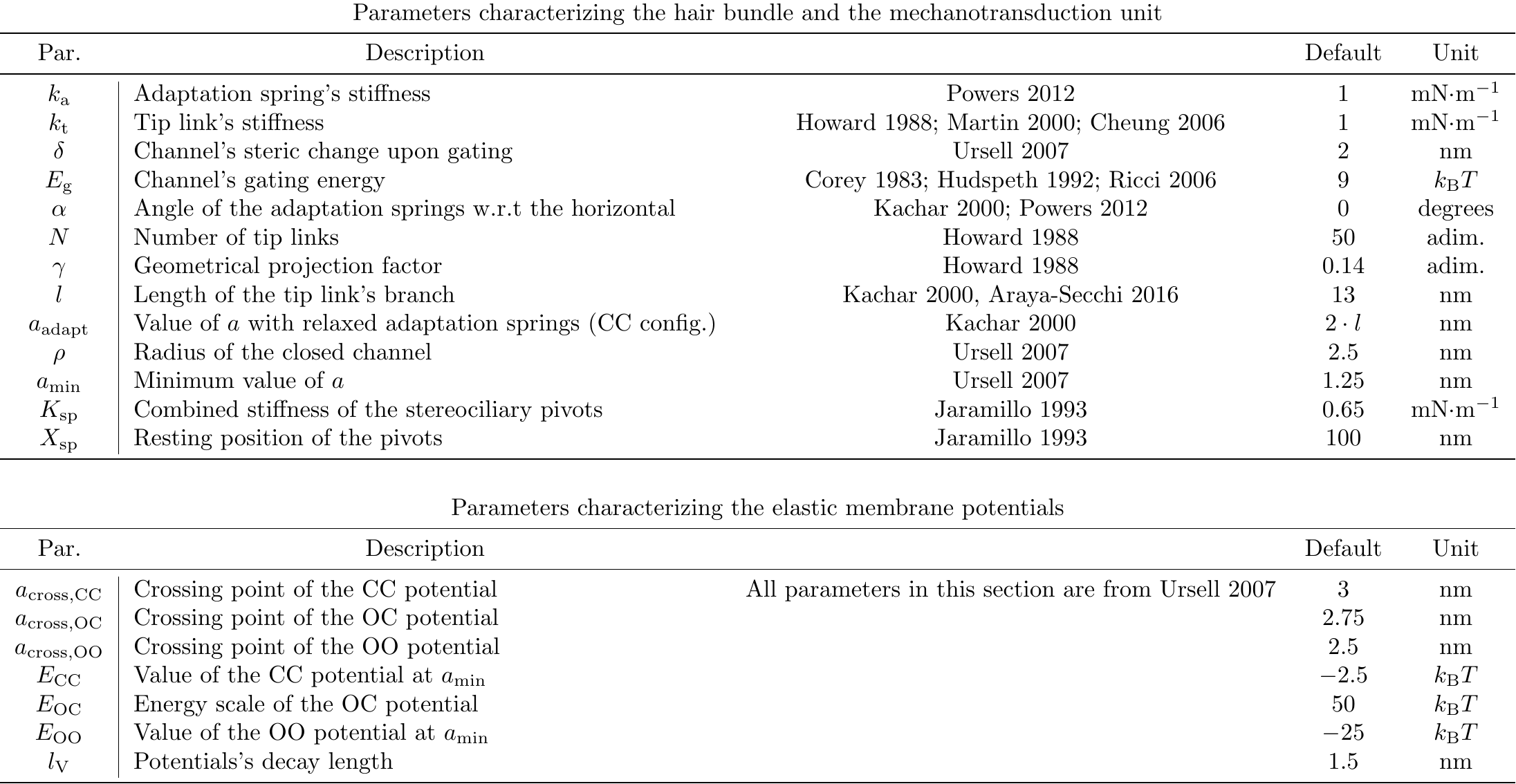}
\label{TableParameters}
\end{table*}

\subsection*{Mathematical Formulation}

We represent schematically our model in Fig.~\ref{FigSchematicModel}. Fig.~\ref{FigSchematicModel}\textit{A} illustrates the geometrical arrangement of a pair of adjacent stereocilia. They have individual pivoting stiffness $k_{\rm SP}$ at their basal insertion points. The displacement coordinate $X$ of the hair bundle's tip along the axis of mechanosensitivity and the coordinate $x$ along the tip link's axis are related by a geometrical factor $\gamma$. With $H$ the height of the tallest stereocilium in the hair bundle and $D$ the distance between its rootlet and that of its neighbor, $\gamma$ is approximately equal to $D/H$~\cite{howard_compliance_1988}. The transduction unit schematized in Fig.~\ref{FigSchematicModel}\textit{A} is represented in more detail in Fig.~\ref{FigSchematicModel} \textit{B} and \textit{C}. In Fig.~\ref{FigSchematicModel}\textit{B}, the stereociliary membrane is orthogonal to the tip link’s central axis. Depending on tip-link tension, the channels are likely to be closed (small tip-link tension, Fig.~\ref{FigSchematicModel}\textit{B}, \textit{Left}) or open (large tip-link tension, Fig.~\ref{FigSchematicModel}\textit{B}, \textit{Right}), and positioned at different locations. The tip link is modeled as a spring of constant stiffness $k_{\rm t}$ and resting length $l_{\rm t}$. It has a current length $x_{\rm t}$ and branches into two rigid strands of length $l$, a distance $d$ away from the membrane. Each strand connects to one MET channel. Due to the global geometry of the hair bundle (Fig.~\ref{FigSchematicModel} \textit{A} and \textit{B}), $x_{\rm t} + d=\gamma(X-X_0)$, where $X_0$ is a reference position of the hair-bundle tip related to the position of the adaptation motors, to which the upper part of the tip link is anchored (see also Hair-Bundle Force and Stiffness and Fig.~S2). The channels' positions are symmetric relative to the tip-link axis, with their attachments to the tip link a distance $2a$ from each other. The channels have cylindrical shapes with axes perpendicular to the membrane plane. They have a diameter $2\rho$ when closed and $2\rho + \delta$ when open, where $\delta$ corresponds to the conformational change of each channel along the membrane plane; we refer to it as the single-channel gating swing. Each tip-link branch inserts a distance $\rho/2$ from the inner edge of each channel. The adaptation springs are parallel to the direction of channel motion. They have stiffness $k_{\rm a}$ and resting length $l_{\rm a}$ and are anchored to two fixed reference positions a distance $L$ away from the tip-link axis.

Under tension, the stereociliary membrane can present different degrees of tenting~\cite{kachar_high-resolution_2000, assad_tip-link_1991}. To account for this geometry, and more generally for the non-zero curvature of the membrane at the tips of stereocilia, we introduce in Fig.~\ref{FigSchematicModel}\textit{C} an angle $\alpha$ between the perpendicular to the tip link’s axis and each of the half membrane planes, along which the channels move. The simpler, flat geometry of Fig.~\ref{FigSchematicModel}\textit{B} is recovered in the case where $\alpha = 0$.
  
The inter-channel forces mediated by the membrane are described by three elastic potentials $V_{{\rm b},n}(a)$, one for each state $n$ of the channel pair (OO, OC, and CC), and are functions of the distance $a$ (Fig.~\ref{FigSchematicModel}D). The index $n$ can be 0, 1, or 2, corresponding to the number of open channels in the transduction unit. We choose analytic expressions and parameters that mimic the shapes of the potentials used to model similar interactions between bacterial MscL channels~\cite{ursell_cooperative_2007, haselwandter_connection_2013} (\textit{Materials and Methods}). In addition to the membrane-mediated elastic force $f_{{\rm b},n} = - {\rm d}V_{{\rm b},n}/{\rm d}a$, force balance on the channels depends on the force $f_{\rm t} = k_{\rm t}(x_{\rm t} - l_{\rm t})$ exerted by the tip link on its two branches and on the force $f_{\rm a} = k_{\rm a}(a_{\rm adapt} - a - n\delta/2)$ exerted by the adaptation springs, where $a_{\rm adapt} = L - l_{\rm a} - 3\rho/2$ is the value of $a$ for which the adaptation springs are relaxed when both channels are closed. Taking into account the geometry and the connection between $x_{\rm t}$ and $X$ given previously, force balance on either of the two channels reads:
\begin{equation}\label{EqForceBalance}
\begin{aligned}
k_{\rm t}[\gamma(X - X_0) - d - l_{\rm t}] = 2\frac{d + a \sin{\alpha}}{a + d \sin{\alpha}}  \times \\ \times \left [k_{\rm a} \left ( a_{\rm adapt} - a -\frac{n}{2}\delta \right ) - \frac{{\rm d}V_{{\rm b},n}(a)}{{\rm d}a} \right ]\, .
\end{aligned}
\end{equation}
In addition, the geometry implies:
\begin{equation}\label{EqGeometry}
d = \sqrt{l^2 - (a \cos{\alpha})^2} - a \sin{\alpha}\, .
\end{equation}
Putting the expressions of $d$ and $V_{{\rm b},n}$ as functions of $a$ into Eq.~\ref{EqForceBalance} allows us to solve for $X$ as a function of $a$, for each state $n$. Inverting these three functions numerically gives three relations $a_n(X)$, which are then used to express all the relevant quantities as functions of the displacement coordinate $X$ of the hair bundle, taking into account the probabilities of the different states. Further details about this procedure are presented in \textit{Materials and Methods}.

Finally, global force balance is imposed at the level of the whole hair bundle, taking into account the pivoting stiffness of the stereocilia at their insertion points into the cuticular plate of the cell (Fig.~\ref{FigTwoChannelModel}\textit{B}, \textit{Inset}, and Fig.~\ref{FigSchematicModel}\textit{A}):
\begin{equation}
F_{\rm ext} = K_{\rm sp}(X - X_{\rm sp}) + F_{\rm t}\, ,
\label{EqBundleForceBalance}
\end{equation}
where $F_{\rm ext}$ is the total external force exerted at the tip of the hair bundle along the $X$ axis, $K_{\rm sp}$ is the combined stiffness of the stereociliary pivots along the same axis, $X_{\rm sp}$ is the position of the hair-bundle tip for which the pivots are at rest, and $F_{\rm t} = N\gamma f_{\rm t}$ is the combined force of the tip links projected onto the $X$ axis, with $N$ being the number of tip links.

In these equations, two related reference positions appear: $X_0$ and $X_{\rm sp}$. As the origin of the $X$ axis is arbitrary, only their difference is relevant. The interpretation of $X_{\rm sp}$ is given just above. As for $X_0$, it sets the amount of tension exerted by the tip links, since the force exerted by the tip link on its two branches reads $f_{\rm t} = k_{\rm t}[\gamma(X - X_0) - d - l_{\rm t}]$. To fix $X_0$---or equivalently the combination $X_0+l_{\rm t}/\gamma$, which appears in this expression---we rely on the experimentally observed hair-bundle movement that occurs when tip links are cut, and which is typically on the order of 100~nm~\cite{assad_tip-link_1991, jaramillo_displacement-clamp_1993}. Therefore, imposing $X = 0$ as the resting position of the hair-bundle tip with intact tip links, Eq.~\ref{EqBundleForceBalance} must be satisfied with $F_{\rm ext} = 0$, $X = 0$, and $X_{\rm sp} = 100$~nm, which formally sets the value of $X_0$ for any predefined $l_{\rm t}$. Solving for $X_0$, however, requires a numerical procedure, the details of which are presented in {\it Materials and Methods}.

All parameters characterizing the system together with their default values are listed in Table~\ref{TableParameters}. The geometrical projection factor $\gamma$ and number of stereocilia $N$ are set, respectively, to 0.14 and 50~\cite{howard_compliance_1988}. The combined stiffness of the stereociliary pivots $K_{\rm sp}$ is set to 0.65~mN$\cdot$m$^{-1}$~\cite{jaramillo_displacement-clamp_1993}. We use a tip-link stiffness $k_{\rm t}$ and an adaptation-spring stiffness $k_{\rm a}$ of 1~mN$\cdot$m$^{-1}$ to obtain a total hair-bundle stiffness in agreement with experimental observations~\cite{howard_compliance_1988}. The length $l$ of the tip-link branch can be estimated by analyzing the structure of protocadherin-15, a protein constituting the tip link’s lower end. Three extracellular cadherin (EC) repeats are present after the kink at the EC8--EC9 interface, which suggests that $l$ is $\sim$12--14~nm~\cite{araya-secchi_elastic_2016}. This estimate agrees with studies based on high-resolution electron microscopy of the tip link~\cite{kachar_high-resolution_2000}. We allow for the branch to fully relax the adaptation springs by choosing $a_{\rm adapt}=2\cdot l$. The parameters $\delta$ and $\rho$ correspond respectively to the amplitude of the conformational change of a single channel in the membrane plane upon gating and to the radius of the closed channel (see Fig.~\ref{FigSchematicModel}\textit{B}). Since the hair-cell MET channel has not yet been crystallized, we rely on the crystal structures of another mechanosensitive protein, the bacterial MscL channel, and choose $\delta=2$~nm and $\rho=2.5$~nm~\cite{ursell_cooperative_2007}. Finally, the channel gating energy $E_{\rm g}$ is estimated in the literature to be on the order of 5--20~$k_{\rm B}T$~\cite{corey_kinetics_1983,hudspeth_hair-bundle_1992, ricci_mechano-electrical_2006}. We use 9~$k_{\rm B}T$ as a default value.

We now focus on the predictions of this model regarding the main biophysical characteristics of hair-bundle mechanics: open probability, force and stiffness as functions of displacement, the twitch during fast adaptation, and effects of Ca\textsuperscript{2+} concentration on hair-bundle mechanics.

\subsection*{Open Probability}

\begin{figure}[h]
\centering
\includegraphics[width=\linewidth]{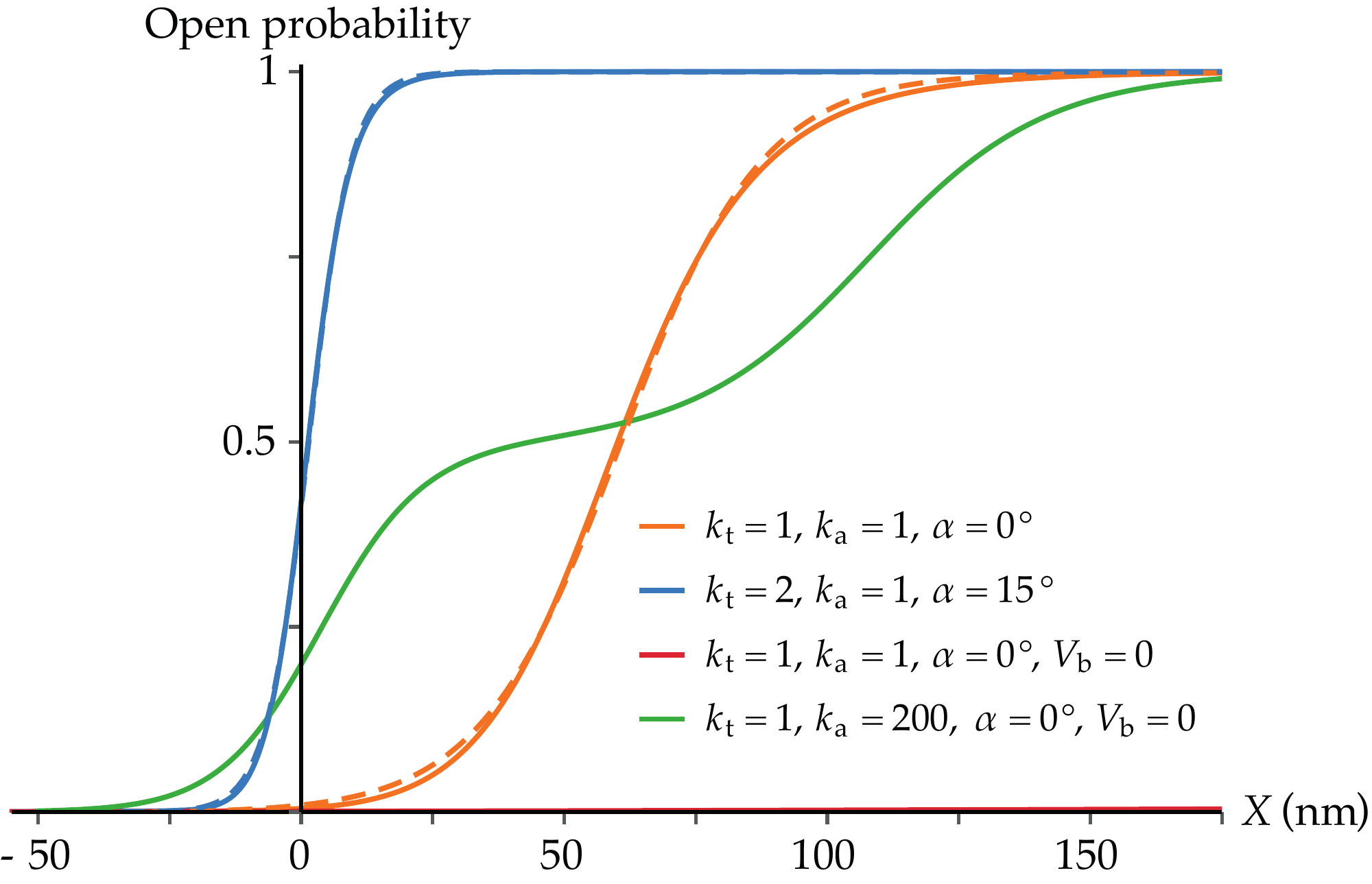}
\caption{Open probability curves as functions of hair-bundle displacement. All curves share a common set of parameters, whose values are specified in Table~\ref{TableParameters}. Parameter values that are not common to all curves are specified below. In addition, for each curve, the value of $X_0$ is set such that the external force $F_{\rm ext}$ applied to the hair bundle vanishes at $X = 0$. (Orange) ($k_{\rm t} = 1$~mN$\cdot$m$^{-1}$, $\alpha = 0 \degree$, $k_{\rm a} = 1$~mN$\cdot$m$^{-1}$, $E_{\rm g} = 9$~$k_{\rm B}T$). The curve is roughly sigmoidal and typical of experimental measurements. (Orange dashed) Fit to a two-state Boltzmann distribution as resulting from the classical gating-spring model, with expression $1/(1 + \exp[z(X_0 - X)/{k_{\rm B}T}])$, where $z \simeq 0.36$~pN, $X_0 \simeq 35$~nm, and $k_{\rm B}T \simeq 4.1$~zJ. Here, $z$ corresponds to the gating force in the framework of the classical gating-spring model. (Blue) ($k_{\rm t} = 2$~mN$\cdot$m$^{-1}$, $\alpha = 15 \degree$, $k_{\rm a} = 1$~mN$\cdot$m$^{-1}$, $E_{\rm g} = 8.8$~$k_{\rm B}T$). The values of $X_0$ and $E_{\rm g}$ have been chosen so that the force is zero at $X = 0$ within the region of negative stiffness, which is required for a spontaneously oscillating hair bundle~\cite{martin_negative_2000}. Channel gating occurs here over a narrower range of hair-bundle displacements. (Blue dashed) Fit to a two-state Boltzmann distribution, with $z \simeq 1.0$~pN, $X_0 \simeq 1.4$~nm, and $k_{\rm B}T \simeq 4.1$~zJ. (Red) ($k_{\rm t} = 1$~mN$\cdot$m$^{-1}$, $\alpha = 0 \degree$, $k_{\rm a} = 1$~mN$\cdot$m$^{-1}$, $E_{\rm g} = 9$~$k_{\rm B}T$, no membrane potentials). The channels remain closed over the whole range of displacements shown in the figure. (Green) ($k_{\rm t} = 1$~mN$\cdot$m$^{-1}$, $\alpha = 0 \degree$, $k_{\rm a} = 200$~mN$\cdot$m$^{-1}$, $E_{\rm g} = 9$~$k_{\rm B}T$, no membrane potentials). The curve presents a plateau around $P_{\rm open} = 0.5$.
}\label{FigOpenprobability}
\end{figure}

To determine the accuracy of the model and to investigate the effect of its parameters, we first focus on the predicted open probability ($P_{\rm open}$) as a function of the hair-bundle displacement $X$, for four sets of parameters (Fig.~\ref{FigOpenprobability}).

For our default parameter set defined above (see also Table~\ref{TableParameters}), the open probability as a function of hair-bundle displacement is a sigmoid that matches the typical curves measured experimentally (orange, continuous curve). It is well fit by a two-state Boltzmann distribution (orange, dashed curve). In this case, the range of displacements over which the channels gate is $\sim$100~nm, in line with experimental measurements~\cite{ricci_mechanisms_2002, he_mechanoelectrical_2004, jia_mechanoelectric_2007, hudspeth_integrating_2014}. Recorded ranges vary, however, from several tens to hundreds of nanometers, depending on whether the hair bundle moves spontaneously or is stimulated, and depending on the method of stimulation (ref.~\cite{meenderink_voltage-mediated_2015} and reviewed in ref.~\cite{fettiplace_physiology_2014}). Increasing the tip-link stiffness~$k_{\rm t}$ and the angle~$\alpha$ compresses this range to a few tens of nanometers (blue curve), matching that measured for spontaneously oscillating hair bundles~\cite{meenderink_voltage-mediated_2015}.

Decreasing the amplitude of the single-channel gating swing~$\delta$ instead broadens the range and shifts it to larger hair-bundle displacements (Fig.~S1). In contrast to the classical model of mechanotransduction, where channel gating is intimately linked to the existence of the single-channel gating swing, here, gating still takes place when $\delta = 0$ due to the membrane elastic potentials. To demonstrate the crucial role played by these potentials, we compare in Fig.~\ref{FigOpenprobability} the open probability curves obtained using the default set of parameters with (orange) and without (red) the bilayer-mediated interaction. Without the membrane contribution, the channels remain closed over the whole range of hair-bundle displacements. The associated curve (red) is barely visible close to the horizontal axis of Fig.~\ref{FigOpenprobability}. It is possible, however, to have the channels gate over this range of displacements without the membrane contribution by choosing a value of $k_{\rm a}$ sufficiently large for the lateral channel motion to be negligible. This configuration mimics the case of immobile channels, as in the classical gating-spring model on timescales that are smaller than the characteristic time of slow adaptation. The resulting curve (green) does not match any experimentally measured open-probability relations: It displays a plateau at $P_{\rm open} = 0.5$ corresponding to the OC state. This state is prevented in the complete model with mobile channels by the membrane-mediated forces. Reintroducing these forces while keeping the same large value of $k_{\rm a}$ hardly changes the open-probability relation, because the channels are maintained too far from each other by the adaptation springs to interact via the membrane. Therefore, we only display one of the two curves here. We illustrate further the influence of the value of $k_{\rm a}$ as well as of the amplitude of the elastic membrane potentials in Fig.~S5.
  
We conclude from these results that our model can reproduce the experimentally observed open-probability relations using only realistic parameters, and that the membrane-mediated interactions as well as the ability of the MET channels to move within the membrane are essential features of the model.

\subsection*{Hair-Bundle Force and Stiffness}

\begin{figure}[h]
\centering
\includegraphics[width=\linewidth]{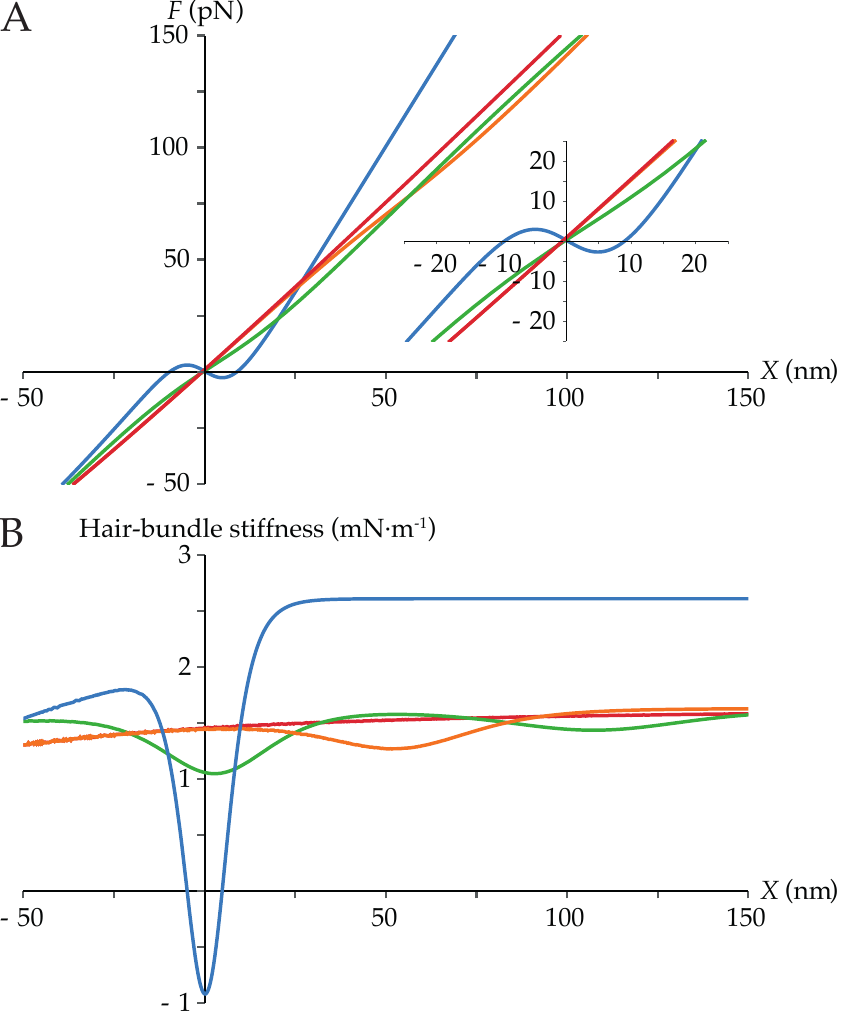}
\caption{Hair-bundle force (\textit{A}) and stiffness (\textit{B}) as functions of hair-bundle displacement. The different sets of parameters are the same as the ones used in Fig.~\ref{FigOpenprobability}, following the same color code. (Orange) The force--displacement curve shows a region of gating compliance, characterized by a decrease in its slope over the gating range of the channels, recovered as a decrease in stiffness over the same range. (Blue) The force--displacement curve shows a region of negative slope, characteristic of a region of mechanical instability. The corresponding stiffness curve shows associated negative values. (Red) Without the membrane elastic potentials, the channels are unable to open and the hair-bundle mechanical properties are roughly linear, except for geometrical nonlinearities. (Green) The curves display two regions of gating compliance, better visible on the stiffness curve.
}\label{FigForceStiffness}
\end{figure}

Two other classical characteristics of hair-cell mechanics are the force- and stiffness- displacement relations. In Fig.~\ref{FigForceStiffness}, we display them using the same sets of parameters and color coding as in Fig.~\ref{FigOpenprobability}.
The predicted forces necessary to move the hair bundle by tens of nanometers are on the order of tens of piconewtons, in line with the literature~\cite{howard_compliance_1988,van_netten_channel_2003}. With our reference set of parameters, the force is weakly nonlinear, associated with a small drop in stiffness (orange curves). When the range of displacements over which the channels gate is sufficiently narrow, a nonmonotonic trend appears in the force, corresponding to a region of negative stiffness (blue curves). In the absence of membrane-mediated interactions---in which case the channels do not gate---the force-displacement curve is nearly linear and the stiffness nearly constant (red curves). The relatively small stiffness variation along the curve is due to the geometry, which imposes a nonlinear relation between hair-bundle displacement and channel motion (Eqs.~\ref{EqForceBalance} and \ref{EqGeometry}). When channel motion is prevented by a large value of $k_{\rm a}$, two separate regions of gating compliance appear, corresponding to the two transitions between the three states (CC, OC, and OO) (green curves). The red curves of this figure demonstrate no contribution from the channels whereas the green curves are again unlike any experimentally measured ones. These results confirm the importance in our model of both the lateral mobility of the channels and the membrane-mediated elastic forces.

We next investigate whether we can reproduce the effects on the force--displacement relation of the slow and fast adaptation~(reviewed in refs.~\cite{eatock_adaptation_2000} and \cite{holt_two_2000}). Slow adaptation is attributed to a change in the position of myosin motors that are connected to the tip link’s upper end and regulate its tension~\cite{howard_compliance_1988,corey_analysis_1983,eatock_adaptation_1987,howard_mechanical_1987,crawford_activation_1989,hacohen_regulation_1989,assad_active_1992,wu_two_1999,kros_reduced_2002}. Here, this phenomenon corresponds to a change in the value of the reference position $X_0$. This parameter affects tip-link tension via the force exerted by the tip link on its two branches: $f_{\rm t} = k_{\rm t}[\gamma(X - X_0) - d - l_{\rm t}]$ (Mathematical Formulation). Starting from the parameters associated with the blue curve and varying $X_0$, we obtain force-displacement relations that are in agreement with experimental measurements (Fig.~S2)~\cite{martin_negative_2000,le_goff_adaptive_2005}.

Fast adaptation is thought to be due to an increase in the gating energy $E_{\rm g}$ of the MET channels, for example, due to Ca\textsuperscript{2+} binding to the channels, which decreases their open probability~\cite{choe_model_1998,cheung_ca2+_2006,ricci_active_2000,wu_two_1999,kennedy_fast_2003}. Starting from the same default curve and changing $E_{\rm g}$ by 1~$k_{\rm B}T$, we obtain a shift in the force--displacement relation (Fig. S3). In this case, the amplitude of displacements over which channel gating occurs remains roughly the same, but the associated values of the external force required to produce these displacements change. Such a shift has been measured in a spontaneously oscillating, weakly slow-adapting cell by triggering acquisition of force--displacement relations after rapid positive or negative steps~\cite{le_goff_adaptive_2005}. During a rapid negative step, the channels close, which we attribute to fast adaptation with an increase in $E_{\rm g}$. In Fig.~S3, increasing $E_{\rm g}$ by 1~$k_{\rm B}T$ increases the value of the force for the same imposed displacement. This mirrors the results in ref.~\cite{le_goff_adaptive_2005}, where a similar outcome is observed when comparing the curve measured after rapid negative steps with that measured after rapid positive steps. From Figs.~S2 and S3, we conclude that our model is capable of reproducing the effects of both slow and fast adaptation on the force--displacement relation.

In summary, the model reproduces realistic force--displacement relations when both lateral channel mobility and membrane-mediated interactions are present. These relations exhibit a region of gating compliance and can even show a region of negative stiffness while keeping all parameters realistic.
  
\subsection*{A Mechanical Correlate of Fast Adaptation, the Twitch}

\begin{figure}[h]
\centering
\includegraphics[width=\linewidth]{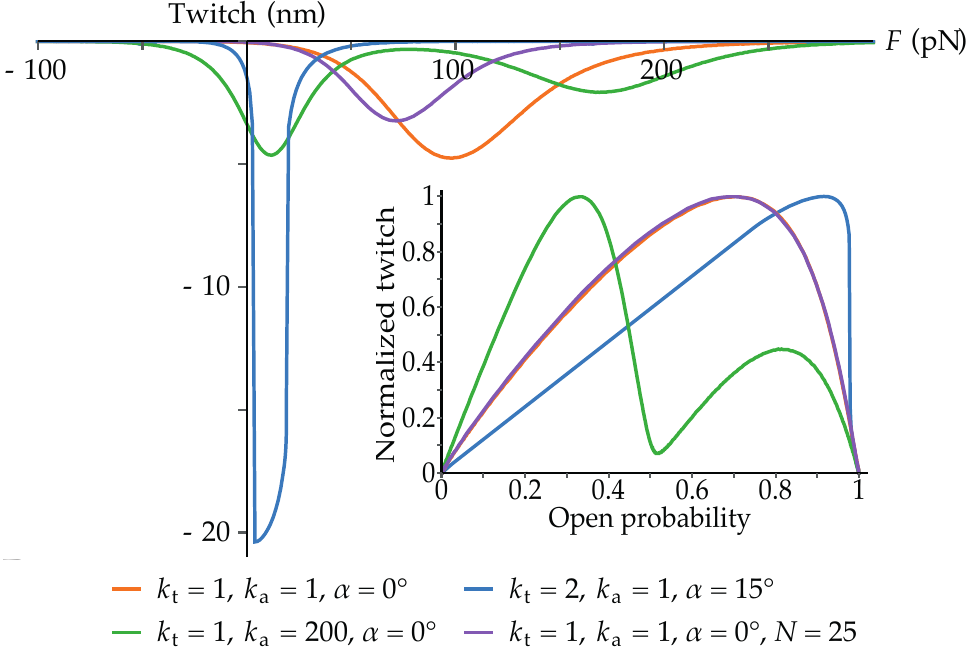}
\caption{Twitch as a function of the external force exerted on the hair bundle (main image) and normalized twitch as a function of the open probability (\textit{Inset}). The different sets of parameters are the same as the ones in Figs.~\ref{FigOpenprobability} and \ref{FigForceStiffness} for the orange, blue and green curves. The additional purple curve is associated with the same parameter set as that of the orange curve, except for the number of intact tip links, set to $N=25$ rather than $N=50$. (Orange) The maximal twitch amplitude for the standard set of parameters is $\sim$5~nm. (Blue) Because of the region of mechanical instability associated with negative stiffness, the corresponding curve for the twitch is discontinuous, as shown by the two regions of near verticality in the blue curve. This corresponds to the two regions of almost straight lines in the normalized twitch. Both of these linear parts are displayed as guides for the eye. (Green) The channels gate independently, producing two distinct maxima of the twitch amplitude. (Purple) The twitch peaks at a smaller force and its amplitude is reduced compared with the orange curve. Plotted as a function of the open probability, however, the two curves are virtually identical.
}\label{FigTwitch}
\end{figure}

Next, we investigate whether our model can reproduce the hair-bundle negative displacement induced by rapid reclosure of the MET channels, known as the twitch~\cite{cheung_ca2+_2006, benser_rapid_1996, ricci_active_2000}. It is a mechanical correlate of fast adaptation, an essential biophysical property of hair cells, which is believed to allow for rapid cycle-by-cycle stimulus amplification~\cite{choe_model_1998}. To reproduce the twitch observed experimentally~\cite{cheung_ca2+_2006, benser_rapid_1996, ricci_active_2000}, we compute the difference in the positions of the hair bundle before and after an increment of $E_{\rm g}$ by 1~$k_{\rm B}T$, and plot it as a function of the external force (Fig. \ref{FigTwitch}).
With the same parameters as in Figs.~\ref{FigOpenprobability} and \ref{FigForceStiffness}, we find twitch amplitudes within the range reported in the literature~\cite{cheung_ca2+_2006, benser_rapid_1996, ricci_active_2000}. They reach their maxima for intermediate, positive forces and drop to zero for large negative or positive forces, as experimentally observed. The twitch is largest and peaks at the smallest force when the hair bundle displays negative stiffness (blue curve), since the channels open then at the smallest displacements. For the green curve, the channels gate independently, producing two distinct maxima of the twitch amplitude, mirroring the biphasic open-probability relation. Note that no curve is shown with the parameter set corresponding to the red curves of Figs.~\ref{FigOpenprobability} and \ref{FigForceStiffness}, since the twitch is nearly nonexistent in that case.

Twitch amplitudes reported in the literature are variable, ranging from $\sim$4~nm in single, isolated hair cells~\cite{cheung_ca2+_2006}, to $>$30~nm in presumably more intact cells within the sensory epithelium~\cite{benser_rapid_1996, ricci_active_2000}. A potential source of variability is the number of intact tip links, since these can be broken during the isolation procedure. We show that decreasing the number of tip links in our model shifts the twitch to smaller forces and decreases its amplitude (Fig.~\ref{FigTwitch}, purple vs. orange curves). Twitch amplitudes are further studied for different values of the adaptation-spring stiffness and amplitudes of the elastic membrane potentials in Fig.~S5. To compare further with experimental data~\cite{benser_rapid_1996, ricci_active_2000}, we also present the twitch amplitude normalized by its maximal value, and plot it as a function of the channels’ open probability (Fig.~\ref{FigTwitch}, \textit{Inset}). The twitch reaches its maximum for an intermediate level of the open probability and drops to zero for smaller or larger values, as measured experimentally~\cite{benser_rapid_1996, ricci_active_2000}.

Another factor that strongly affects both the amplitude and force dependence of the twitch is the length $l$ of the tip-link branching fork. For a long time, the channels were suspected to be located at the tip link's upper end, where the tip-link branches appear much longer~\cite{kachar_high-resolution_2000}. With long branches, the twitch is tiny and peaks at forces that are too large (Fig.~S4), unlike what is experimentally measured. This observation provides a potential physiological reason why the channels are located at the tip link's lower end rather than at the upper end as previously assumed~\cite{zhao_elusive_2015,spinelli_bottoms_2009}. There are two more reasons why our model requires the channels to be located at the lower end of the tip link. First, as shown in Figs.~\ref{FigOpenprobability}--\ref{FigTwitch}, some degree of membrane tenting increases the sensitivity and nonlinearity of the system, as well as the amplitude of the twitch. While it is straightforward to obtain the necessary membrane curvature at the tip of a stereocilium, this is not the case on its side. Second, while pulling on the channels located at the tip compels them to move toward one another, doing so with the channels located on the side would instead make them slide down the stereocilium, impairing the efficiency of the mechanism proposed in this work.

In summary, our model reproduces correctly the hair-bundle twitch as well as its dependence on several key parameters. It therefore includes the mechanism that can mediate the cycle-by-cycle sound amplification by hair cells.

\subsection*{Effect of Ca\textsuperscript{2+} Concentration on Hair-Bundle Mechanics}
  
Our model can also explain the following important results that have so far evaded explanation. First, it is established that, with increasing Ca\textsuperscript{2+} concentration, the receptor current vs. displacement curve shifts to more positive displacements, while its slope decreases~\cite{corey_kinetics_1983}. Second, within the framework of the classical gating-spring model, Ca\textsuperscript{2+} concentration appears to affect the magnitude of the gating swing~\cite{tinevez_unifying_2007}: When a hair bundle is exposed to a low, physiological, Ca\textsuperscript{2+} concentration of 0.25~mM, the force--displacement relation presents a pronounced region of negative slope, and the estimated gating swing is large, on the order of 9--10~nm. But when the same hair bundle is exposed to a high Ca\textsuperscript{2+} concentration of $\sim$1~mM, the region of negative stiffness disappears, and the estimated gating swing becomes only half as large.

In our model, it is the decrease of the interchannel distance following channel opening that, transmitted onto the tip link's main axis, effectively plays the role of the classical gating swing (Figs.~\ref{FigTwoChannelModel} and \ref{FigSchematicModel} and Movie~S1). To quantify the change of tip-link extension as the channels open, we introduce a new quantity, which we call the gating-associated tip-link extension (GATE). It is defined mathematically as $d_{\rm OO}-d_{\rm CC}$, where $d_{\rm OO}$ and $d_{\rm CC}$ are the respective values of the distance $d$ in the OO and CC states. To study the influence of Ca\textsuperscript{2+} concentration on the GATE, we hypothesize that Ca\textsuperscript{2+} ions favor the closed conformation of the channels over the open one, that is, that the energy difference $E_{\rm g}$ between the two states increases with Ca\textsuperscript{2+} concentration~\cite{choe_model_1998, cheung_ca2+_2006}.
\begin{figure}[h]
\centering
\includegraphics[width=\linewidth]{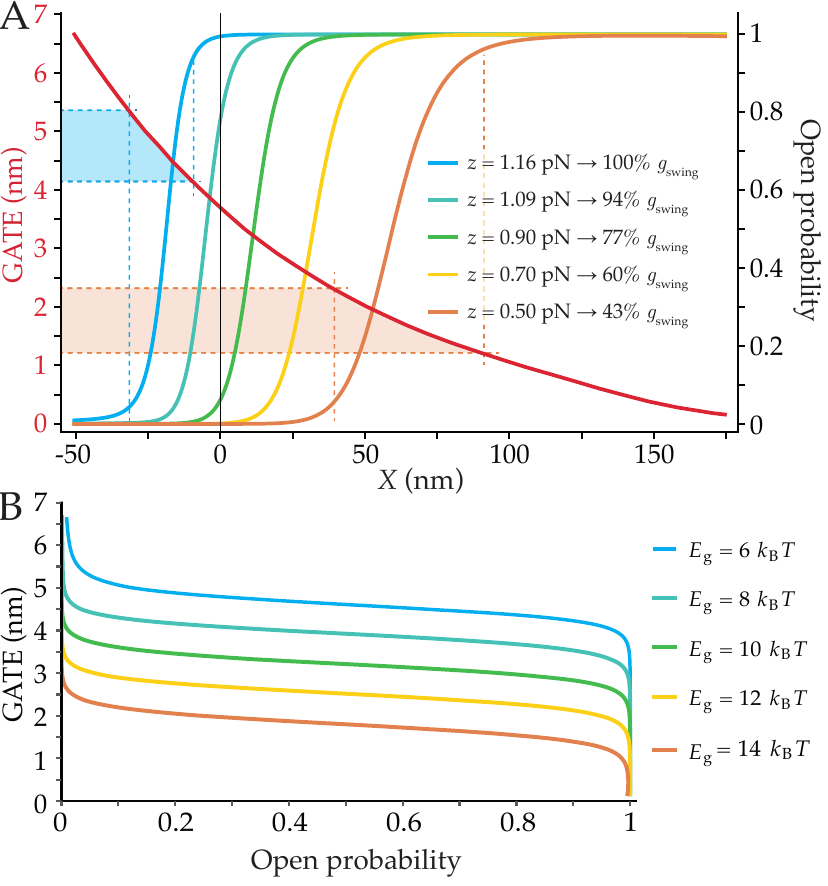}
\caption{GATE and open probability as functions of hair-bundle displacement (\textit{A}) and GATE as a function of the open probability (\textit{B}), for different values of the channel gating energy $E_{\rm g}$. (\textit{A}) Open-probability curves are generated by using the default parameter set of the blue curves of Figs.~\ref{FigOpenprobability}--\ref{FigTwitch}, and otherwise different values of the channel gating energy $E_{\rm g}$, as indicated directly on \textit{B}. The GATE as a function of $X$ (red curve) depends only on the geometry of the system, such that only one curve appears here. We indicate, in addition, directly on the image the values of the single-channel gating force $z$ obtained by fitting each open-probability relation with a two-state Boltzmann distribution, as done in Fig.~\ref{FigOpenprobability} for the orange and blue curves. We report, together with these values, the relative magnitudes of the single-channel gating swing $g_{\rm swing}$ obtained with the classical model. (\textit{B}) The GATE values are plotted as functions of the open probability, for each chosen value of $E_{\rm g}$.
}\label{FigGATE}
\end{figure}
Within this framework, we expect to see the following effect of Ca\textsuperscript{2+} on the GATE, via the change of $E_{\rm g}$: Higher Ca\textsuperscript{2+} concentrations correspond to higher gating energies, causing the channels to open at greater positive hair-bundle displacements. Greater displacements in turn correspond to smaller values of the inter-channel distance before channel opening. Since the final position of the open channels is always the same (at $a=a_{\rm min}$, where the OO membrane potential is minimum; Fig.~\ref{FigSchematicModel}\textit{D}), the change of the interchannel distance induced by channel opening is smaller for higher Ca\textsuperscript{2+} concentrations. As a result, the GATE experienced by the tip link is smaller for higher Ca\textsuperscript{2+} concentrations, in agreement with the experimental findings cited above.

We study this effect quantitatively in Fig.~\ref{FigGATE}. In Fig.~\ref{FigGATE}\textit{A}, we plot simultaneously the GATE and $P_{\rm open}$ as functions of the hair-bundle displacement $X$, for five values of $E_{\rm g}$. Although the function ${\rm GATE}(X)$ spans the whole range of displacements, the relevant magnitudes of the GATE are constrained by the displacements for which channel opening is likely to happen; we use as a criterion that $P_{\rm open}$ must be between 0.05 and 0.95. The corresponding range of displacements depends on the position of the $P_{\rm open}$ curve along the horizontal axis, which ultimately depends on $E_{\rm g}$. We display in Fig.~\ref{FigGATE}\textit{A} the two ranges of hair-bundle displacements (dashed vertical lines) associated with the smallest and largest values of $E_{\rm g}$, together with the amplitudes of the GATE within these intervals (dashed horizontal lines). For $E_{\rm g} = 6$~$k_{\rm B}T$ (blue curve and GATE interval), the size of the GATE is on the order of $4.1\text{--}5.3$~nm, whereas it is on the order of $1.3\text{--}2.3$~nm for $E_{\rm g} = 14$~$k_{\rm B}T$ (orange curve and GATE interval). In general, larger values of the channel gating energy $E_{\rm g}$ cause smaller values of the GATE.

To compare directly with previous analyses, we next fit the open-probability relations of Fig.~\ref{FigGATE}\textit{A} with the gating-spring model, obtaining the corresponding single-channel gating forces $z$. This procedure allows us to quantify the change of the magnitude of an effective gating swing $g_{\rm swing}$ with $E_{\rm g}$ by the formula $z=g_{\rm swing}k_{\rm gs}\gamma$, where $k_{\rm gs}$ is the stiffness of the gating spring. We give directly on the panel the relative values of $g_{\rm swing}$ obtained by this procedure. Taking, for example, $k_{\rm gs}=1$~mN$\cdot$m$^{-1}$, $g_{\rm swing}$ ranges from 8.3~nm for $E_{\rm g} = 6$~$k_{\rm B}T$ to 3.6~nm for $E_{\rm g} = 14$~$k_{\rm B}T$.

In Fig.~\ref{FigGATE}\textit{B}, we show the GATE as a function of $P_{\rm open}$ for the different values of $E_{\rm g}$. For each curve, the amplitude of the GATE is a decreasing function of $P_{\rm open}$ that presents a broad region of relatively weak dependence for most $P_{\rm open}$ values. These results demonstrate that the GATE defined within our model decreases with increasing values of $E_{\rm g}$, corresponding to increasing Ca\textsuperscript{2+} concentrations. In addition, the same dependence is observed for the effective gating swing estimated from fitting the classical gating-spring model to our results, as it is when fit to experimental data~\cite{tinevez_unifying_2007}.

Finally, we can see from Fig.~\ref{FigGATE}\textit{A} that the predicted open-probability vs. displacement curves shift to the right and their slopes decrease with increasing values of $E_{\rm g}$, a behavior in agreement with experimental data (see above and ref.~\cite{corey_kinetics_1983}). Together with the decrease in the slope, the region of negative stiffness becomes narrower (Fig.~S3) and even disappears for a sufficiently large value of $E_{\rm g}$ (Fig.~S3, yellow curve). This weakening of the gating compliance has been measured in hair bundles exposed to a high Ca\textsuperscript{2+} concentration~\cite{tinevez_unifying_2007}.

In summary, our model explains the shift in the force-displacement curve as well as the changes of the effective gating swing and stiffness as functions of Ca\textsuperscript{2+} concentration.

\section*{Discussion}

We have designed and analyzed a two-channel, cooperative model of hair-cell mechanotransduction. The proposed geometry includes two MET channels connected to one tip link. The channels can move relative to each other within the stereociliary membrane and interact via its induced deformations, which depend on whether the channels are open or closed. This cross-talk produces cooperative gating between the two channels, a key feature of our model. Most importantly, because the elastic membrane potentials are affected by channel gating on length scales larger than the proteins’ conformational rearrangements, and because the channels can move in the membrane over distances greater than their own size, the model generates an appropriately large effective gating swing without invoking unrealistically large conformational changes. Moreover, even when the single-channel gating swing vanishes, the effective gating swing determined by fitting the classical model to our results does not. In this case, the conformational change of the channel is orthogonal to the membrane plane and its gating is triggered only by the difference in membrane energies between the OO and CC states. We have shown that our model reproduces the hair bundle’s characteristic current-- and force--displacement relations as well as the existence and characteristics of the twitch, the mechanical correlate of fast adaptation. It also explains the puzzling effects of the extracellular Ca\textsuperscript{2+} concentration on the magnitude of the estimated gating swing and on the spread of the negative-stiffness region, features that are not explained by the classical gating-spring model.

In addition to reproducing these classical features of hair-cell mechanotransduction, our model may be able to account for other phenomena that have had so far no---or only unsatisfactory---explanations. One of them is the flick, a small, voltage-driven hair-bundle motion that requires intact tip links but does not rely on channel gating~\cite{cheung_ca2+_2006, ricci_active_2000, meenderink_voltage-mediated_2015}. It is known that changes in membrane voltage modulate the membrane mechanical tension and potentially the membrane shape by changing the interlipid distance~\cite{zhang_voltage-induced_2001, breneman_hair_2009}, but it is not clear how this property can produce the flick. This effect could be explained within our framework as a result of a change in the positions of the channels following the change in interlipid distance driven by voltage. This would in turn change the extension of the tip link and thus cause a hair-bundle motion corresponding to the flick.

Another puzzling observation from the experimental literature is the recordings of transduction currents that appear as single events but with conductances twofold to fourfold that of a single MET channel~\cite{pan_tmc1_2013, beurg_conductance_2014}. Because tip-link lower ends were occasionally observed to branch into three or four strands at the membrane insertion~\cite{kachar_high-resolution_2000}, one tip link could occasionally be connected to as many channels. According to our model, these large-conductance events could therefore reflect the cooperative openings of coupled channels. 

Our model predicts that changing the membrane properties must affect the interaction between the MET channels, potentially disrupting their cooperativity and in turn impairing the ear’s sensitivity and frequency selectivity. For example, if the bare bilayer thickness were to match more closely the hydrophobic thickness of the open state of the channel rather than that of the closed state, the whole shape of the elastic membrane potentials would be different. In such a case, the open probability vs. displacement curves would be strongly affected, and gating compliance and fast adaptation would be compromised. Potentially along these lines, it was observed that chemically removing long-chain---but not short-chain---phospholipid PiP2 blocked fast adaptation~\cite{hirono_hair_2004}. With a larger change of membrane thickness, one could even imagine reversing the roles of the OO and CC membrane-mediated interactions. This would potentially change the direction of fast adaptation, producing an “anti-twitch”, a positive hair-bundle movement due to channel reclosure. Such a movement has indeed been measured in rat outer hair cells~\cite{kennedy_force_2005}. Whether it was produced by this or a different mechanism remains to be investigated.

Our model fundamentally relies on the hydrophobic mismatch between the MET channels and the lipid bilayer. Several studies have demonstrated that the lipids with the greatest hydrophobic mismatch with a given transmembrane protein are depleted from the protein's surrounding. The timescale of this process is on the order of a 100~ns for the first shell of annular lipids~\cite{beaven_gramicidin_2017}. It is much shorter than the timescales of MET-channel gating and fast adaptation. Therefore, it is possible that lipid rearrangement around a MET channel reduces the hydrophobic mismatch and thus decreases the energy cost of the elastic membrane deformations, lowering in turn the importance of the membrane-mediated interactions in hair-bundle mechanics. However, such lipid demixing in the fluid phase of a binary mixture is only partial, on the order of 5--10\%~\cite{yin_hydrophobic_2012}. Furthermore, ion channels are known to bind preferentially specific phospholipids such as PiP2~\cite{suh_pip2_2008}, further suggesting that the lipid composition around a MET channel does not vary substantially on short timescales. We therefore expect the effect of this fast lipid mobility to be relatively minor. Slow, biochemical changes of the bilayer composition around the channels, however, could have a stronger effect. It would be interesting for future studies to investigate the role played by lipid composition around a MET channel on its gating properties and how changes in this composition affect hair-cell mechanotransduction.

\matmethods{

\subsection*{Membrane-Mediated Interaction Potentials} 

The one-dimensional interaction potentials mediated by the membrane between two mechanosensitive channels of large conductance (MscL) in \textit{Escherichia coli} have been modeled by Ursell {\it et al.}~\cite{ursell_cooperative_2007, phillips_emerging_2009}. Here, we mimic the shape of the potentials used in that study with the following analytic expressions: 
\begin{equation}
{\footnotesize  \begin{aligned}
V_{{\rm b},0}(a) = & E_{\rm CC} \left ( \frac{a - a_{\rm cross, CC}}{a_{\rm min} - a_{\rm cross, CC}} \right ) \exp \left [ - \left ( \frac{a - a_{\rm min}}{l_{\rm V}} \right)^2 \right]\\
V_{{\rm b},1}(a) = & E_{\rm OC} \left [ \frac{(a_{\rm cross,OC} - a) (a_{\rm cross, OC} - a_{\rm min})^2}{(a - a_{\rm min})^3} \right ] \times\\
& \times \exp \left [ - \left ( \frac{a - a_{\rm min}}{l_{\rm V}} \right)^2 \right]\\
V_{{\rm b},2}(a) = & E_{\rm CC} \left ( \frac{a - a_{\rm cross, OO}}{a_{\rm min} - a_{\rm cross, OO}} \right ) \exp \left [ - \left ( \frac{a - a_{\rm min}}{l_{\rm V}} \right)^2 \right]\, ,
\end{aligned}}
\end{equation}
where the coordinate $a$ corresponds to the distance between either of the two anchoring points of the tip link and the tip link's central axis (Fig.~\ref{FigSchematicModel}). Note that this choice of coordinate is different from that of Ursell {\it et al.}~\cite{ursell_cooperative_2007}, who chose to represent their potentials as functions of the channels’ centre-to-centre distance. The different parameters entering these expressions, with their associated numerical values used to generate the results presented in this work, are as follows: $a_{\rm min} = 1.25$~nm represents the minimal value reached by the variable $a$; $l_{\rm V} = 1.5$~nm is the characteristic length over which the membrane-mediated interaction decays; $a_{\rm cross, CC} = 3$~nm, $a_{\rm cross,OC} = 2.75$~nm, and $a_{\rm cross,OO} = 2.5$~nm are the respective values of the variable $a$ for which the membrane potentials $V_{{\rm b},0}$, $V_{{\rm b},1}$, and $V_{{\rm b},2}$ cross the $X = 0$ axis; $E_{\rm CC} = -2.5$ $k_{\rm B}T$ and $E_{\rm OO} = -25$ $k_{\rm B}T$ represent, respectively, the values of the potentials $V_{{\rm b},0}$ and $V_{{\rm b},2}$ at $a = a_{\rm min}$; and finally, $E_{\rm OC} = 50$ $k_{\rm B}T$ is an energy scale that describes the global amplitude of $V_{{\rm b},1}$. A graphical representation of the resulting elastic potentials is shown in Fig.~\ref{FigSchematicModel}\textit{D}.

\subsection*{Open Probability}

The open probability of the channels ($P_{\rm open}$) depends on a total energy that is the sum of the following contributions: the elastic energy of the two adaptation springs $E_{{\rm a},n} = 2 \cdot k_{\rm a} ({a}_{\rm adapt} - a - n\delta/2)^2/2$ (for $a < {a}_{\rm adapt} - n\delta/2$, zero otherwise), the elastic energy of the tip link $E_{{\rm t},n} = k_{\rm t}(\gamma(X-X_0) - d - l_{\rm t})^2/2$ (for $\gamma(X-X_0) > d + l_{\rm t}$, zero otherwise), the membrane mechanical energy $V_{{\rm b},n}(a)$ detailed above, and the energy due to channel gating $n \times E_{\rm g}$, where $E_{\rm g}$ is the gating energy of a single channel. Adding these four contributions and using the computed relations $a_n(X)$ (see Numerical Solution of the Model for further details), one can compute the total energy $E_{{\rm tot},n}(X)$ associated with each channel configuration $n$ at every displacement $X$ of the hair bundle. The probability weights of the different channel states are, respectively, $W_{\rm CC} = \exp(- E_{{\rm tot},0}(X))$, $W_{\rm OC} = 2\exp(- E_{{\rm tot},1}(X))$, and $W_{\rm OO} = \exp(- E_{{\rm tot},2}(X))$, the factor two in $W_{\rm OC}$ reflecting the fact that the OC state comprises two canonical states, open-closed and closed-open. Furthermore, the probability of each configuration is equal to its associated probability weight divided by the sum $W_{\rm OO} + W_{\rm OC} + W_{\rm CC}$. At the level of the whole hair bundle, and under the hypothesis that all channel pairs are identical, the overall open probability of the channels finally reads: $P_{\rm open} = P_{\rm OO} + P_{\rm OC}/2$.

\subsection*{Numerical Solution of the Model}

To compute the model outcomes---including the open probability as discussed above---we first need to solve Eq.~\ref{EqForceBalance} to find the three relations $a_n(X)$. This equation, however, cannot be solved analytically for $a$. To obtain numerical solutions, we first solve it analytically for $X$ and obtain three expressions for $X(a,n)$. We then produce a set of three tables of numerical values $X_{n,i} = X(a_i, n)$, where $a_i = a_{\rm min} + i \cdot \Delta a$ is a set of values of the coordinate $a$, equispaced by a length $\Delta a$. To produce tables $a_{n,j} = a_n(X_j)$ that share a common set of entries $X_j$, we first generate a table of entries for the variable $X$, regularly spaced: $X_j = X_{\rm min} + j \cdot \Delta X$. For each entry $X_j$, we then take in the original table $X_{n,i} = X(a_i,n)$ the value $X_{n,i}$ that is the closest to $X_j$. We then choose for $a_{n,j}$ the corresponding value $a_i$. As a result, we obtain three stepwise functions $a_n(X)$ such that $a_n(X) = a_{n,j}$ for all $X$ in $[X_j, X_{j+1}[$. We further smoothen these functions by interpolating linearly the values of $a$ between two neighboring plateaus, to avoid discontinuities and obtain a continuous, piecewise-linear function.

\subsection*{Reference Tip-Link Tension}

As described in Mathematical Formulation, the parameter $X_0$, which sets the reference tension in the tip link, is determined by imposing the force-balance Eq.~\ref{EqBundleForceBalance} with $F_{\rm ext} = 0$~N, $X = 0$~m, and $X_{\rm sp} = 100$~nm. To impose this condition, however, one needs the global tip-link force $F_{\rm t} = N \gamma f_{\rm t}$, which itself depends on $X_0$ via the respective probabilities of the three channel states (OO, OC and CC). In addition, as discussed above, the curves $a_n(X)$ can only be computed numerically, meaning that no closed analytic expression can be obtained for $X_0$. We therefore proceed numerically according to the following scheme: We first generate a table of numerical values of $X_0$ and compute the associated table of tip-link forces $f_{\rm t}$ that satisfy force balance at the level of the whole hair bundle (namely, that solve Eq.~\ref{EqBundleForceBalance} with $F_{\rm ext} = 0$~N, $X = 0$~m, and $X_{\rm sp} = 100$~nm). We then insert these values into the force-balance condition at the level of the individual MET channels (Eq. \ref{EqForceBalance}), evaluated at $X = 0$. The proper value for $X_0$ corresponds to the element for which this condition is satisfied. This ensures force balance both at the level of the individual MET channels and at the level of the whole hair bundle. This procedure corresponds to determining the reference tension in the tip links for a given hair-bundle displacement.

}

\acknow{We thank the members of the A.S.K. laboratory for comments on the manuscript. Work on this project in the A.S.K. laboratory was supported by The Royal Society Grant $RG140650$, Wellcome Trust Grant $108034/Z/15/Z$, and the Imperial College Network of Excellence Award. T.R. was supported by the LabEx CelTisPhyBio ANR-10-LABX-0038.}

\showmatmethods{} 
  
\showacknow{} 
  

\begin{widetext}

\includegraphics[width=\linewidth]{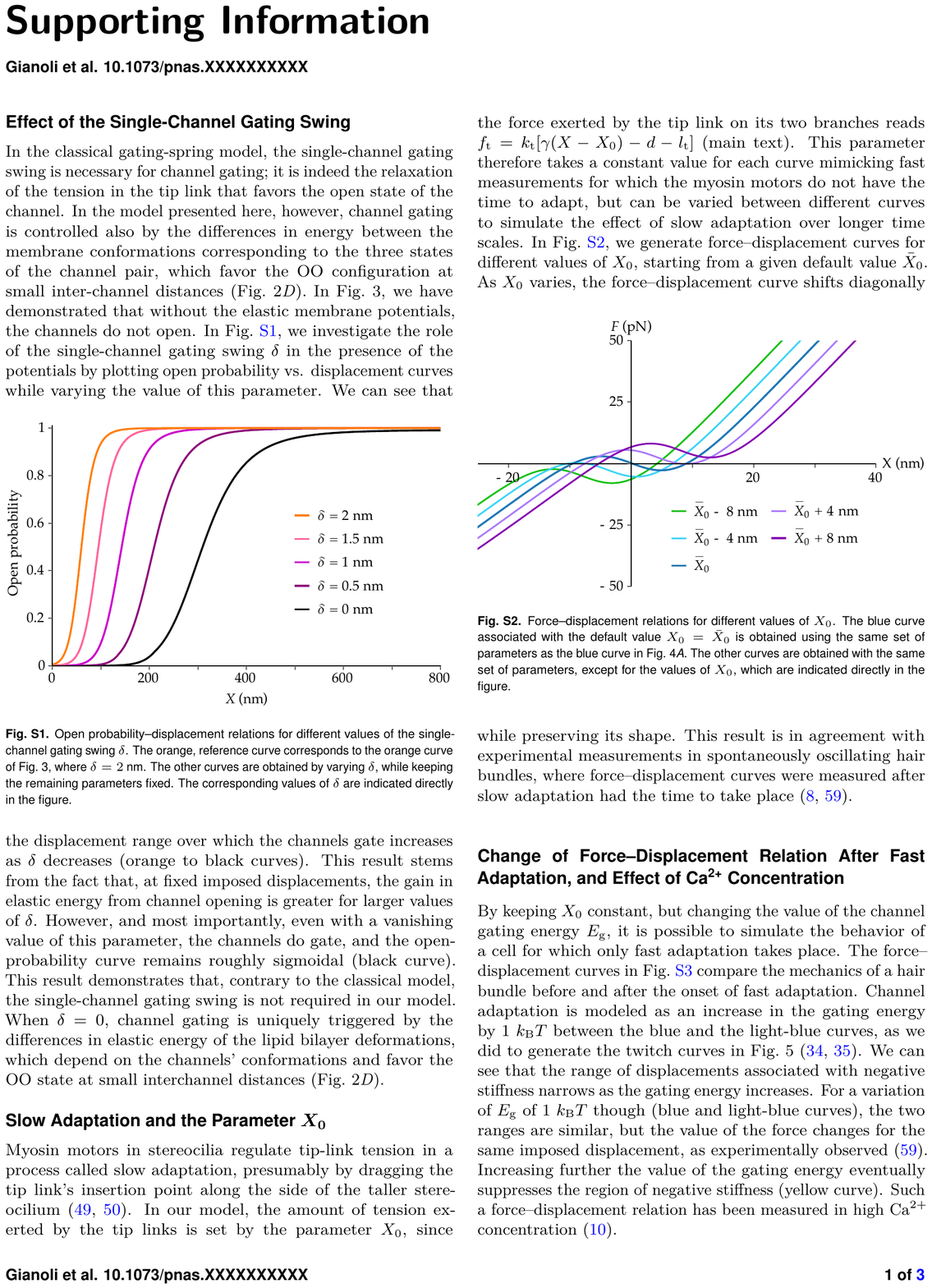}
\includegraphics[width=\linewidth]{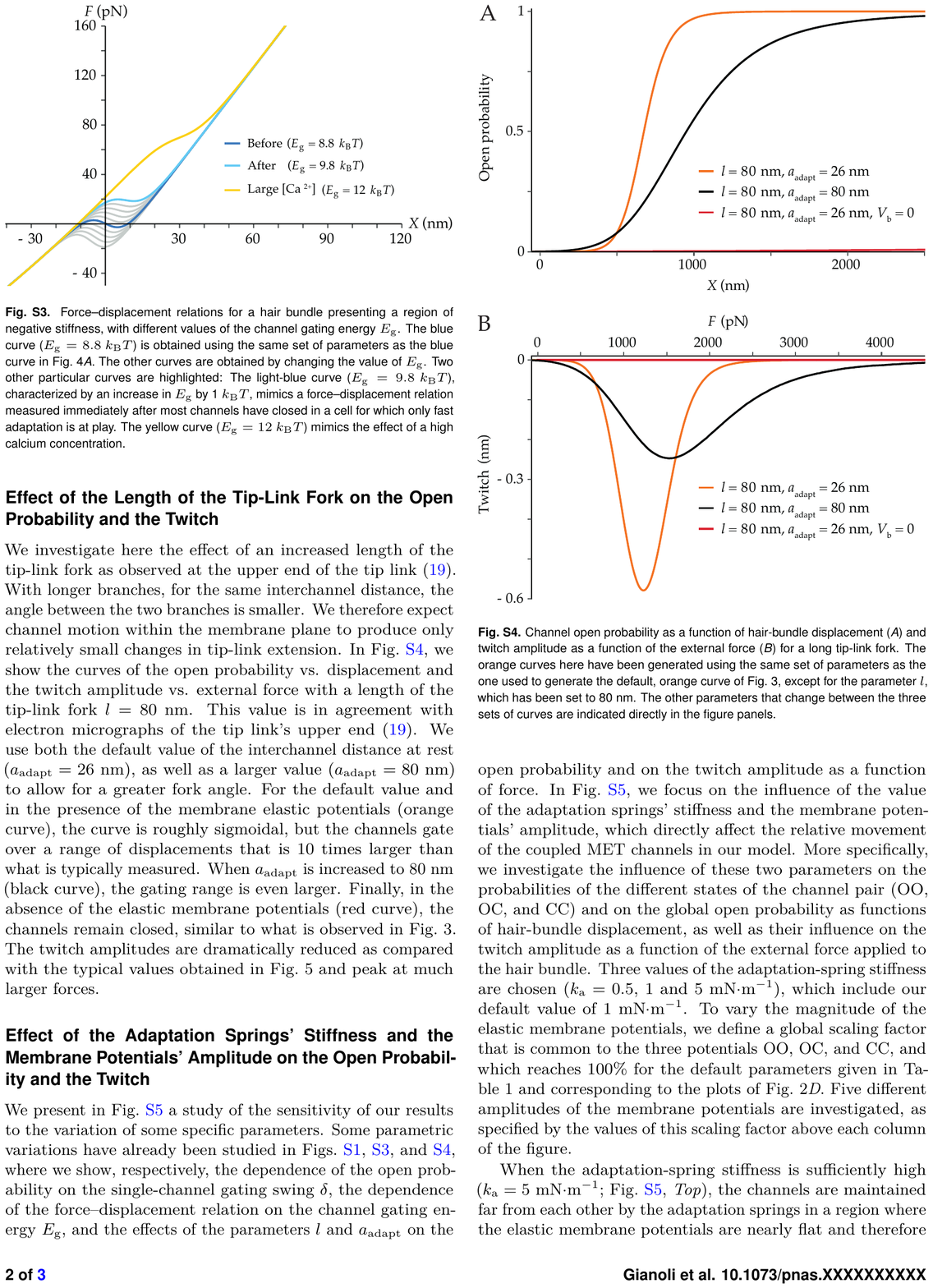}
\includegraphics[width=\linewidth]{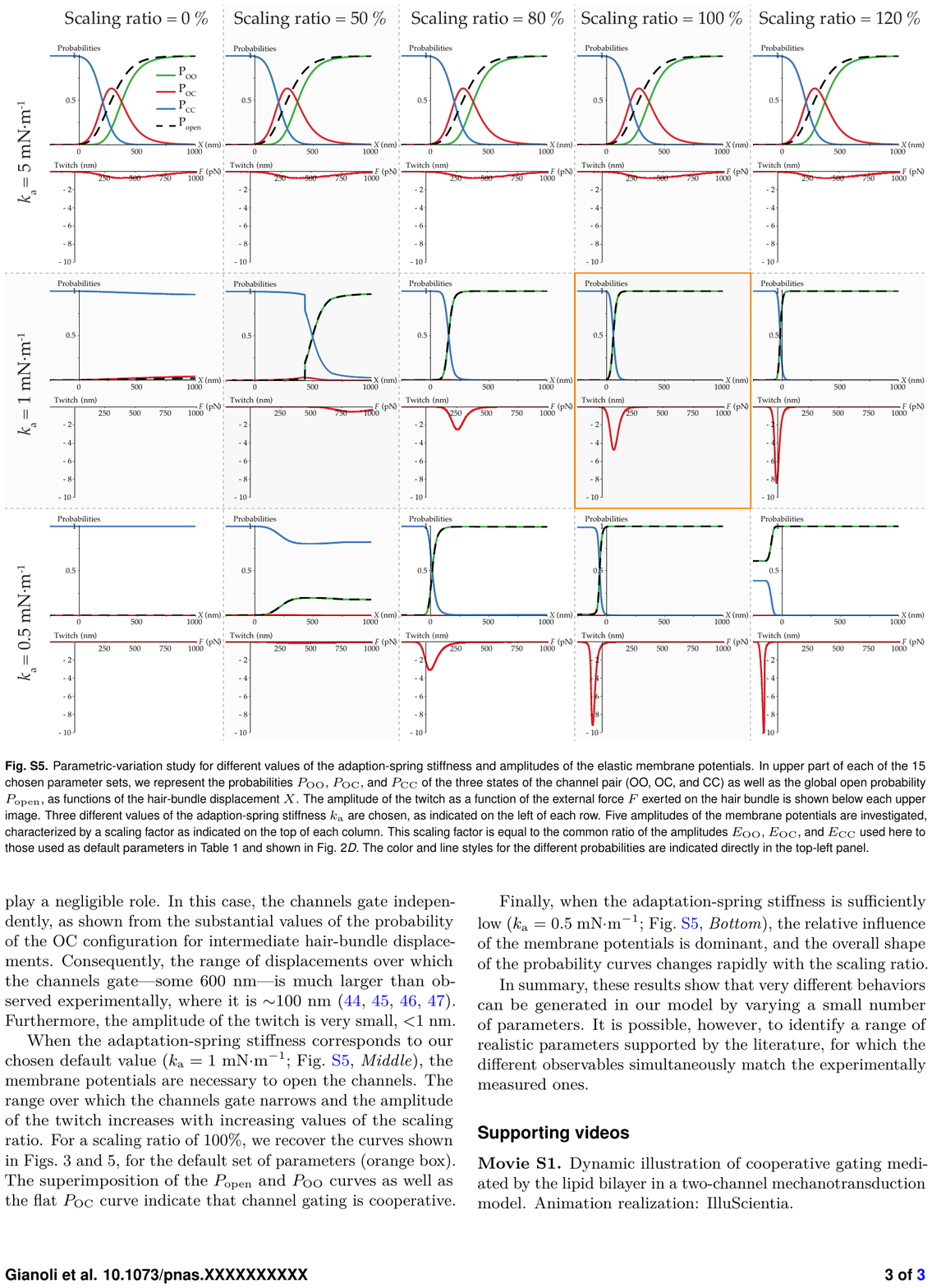}

\end{widetext}

\end{document}